\documentclass{article}
\usepackage{authblk}
\usepackage{amsthm}
\usepackage{amsmath,bm}
\usepackage{enumerate}
\usepackage{amssymb}
\usepackage{latexsym}
\usepackage{amsfonts}
\usepackage{color}
\usepackage{secdot}
\usepackage{mathrsfs}
\usepackage{epsfig}

\usepackage{hyperref}
\usepackage{graphicx}
\usepackage{subcaption}
\usepackage[export]{adjustbox}
\usepackage{wrapfig}
\usepackage[rightcaption]{sidecap}
\usepackage{enumerate}

\usepackage{play}
\usepackage{epsfig}

\title{Permutation extropy: a time series complexity measure}
\author[1,2]{Ritik Roshan Giri}
\author[2]{Suchandan Kayal}
\affil[1]{Department of Mathemtics and Statistics, Indian Institute of Technology Kanpur, India}
\affil[2]{Department of Mathematics, National Institute of Technology Rourkela, India }

\begin{document}
\maketitle
\maketitle
\begin{abstract}

On account of a greater need for understanding the complexity of time series like physiological time series, financial time series, and many more that enter into picture for their inculpation with real-world problems, several complexity parameters have already been proposed in the literature. Permutation entropy, Lyapunov exponents are such complexity parameters out of many. In this article, we introduce a new time series complexity parameter, that is, the permutation extropy. The failure of permutation entropy in correctly specifying complexity of some chaotic time series motivates us to come up with a better complexity parameter, hence we propose this permutation extropy measure. We try to combine the ideas behind the permutation entropy and extopy to construct this measure. We also validate our proposed measure using several chaotic maps like logistic map, Henon map and Burger map. We apply the proposed complexity parameter to study the complexity of financial time series of the stock market and time series constructed using WHO data, finding a better complexity specification than permutation entropy. The proposed measure is kind of robust, fast calculation and invariant with respect to monotonous nonlinear transformation like permutation entropy, but it gives us a better result in specifying complexity in some cases.

\vspace{0.5cm}

{\textbf{Keywords: }}{ Complexity, Permutation extropy, Logistic map, Henon map, Burger map}
\end{abstract}

\section{Introduction}

Recently, the efficient measurement of the complexity of different dynamical systems, such as nonlinear time series, has gained a lot of attention from researchers working in the area related to information theory and applied statistics. Several measures based on information theory are developed which are useful to estimate the complexity and different behaviors of time series such as periodic and chaotic behavior efficiently, e.g. entropy [1-5], fractal dimensions [6], and Lyapunov exponents [7,8].  Out of those various complexity measures, entropy has become widely recognized which is basically the measurement of uncertainty and degree of randomness of the dynamical system. The measurement of uncertainty plays an important role in the prediction of the complexity of a system. Shannon (1948) proposed an information-theoretic approach to entropy, which is related to the uncertainty of a random variable. Consider a discrete time series $\{X_t\}, t\geq0$ that takes values up to time level $t$ among $\{x_1,\ldots, x_t\}$ with respective probabilities $\{p_{1},\ldots, p_{t}\}$, then the measurement of uncertainty/entropy proposed by Shannon is defined as:

\begin{equation}\label{eqn 1}
H(X_{t}) = -\sum_{i=1}^{t} p_{i}\log p_{i},
\end{equation}
where $\log(.)$ denotes the natural logarithm, $p_{i}=P(X=x_{i})$. This equation basically provides information about the uncertainty involved with the value that the time series will take at the time level $t+1$. For calculating the complexity of time series that take infinitely many numbers of values Bandt and Pompe [4] introduced the concept of permutation entropy based on the usual Shanon entropy. The permutation entropy value of a time series can be calculated by comparing the neighbouring values of each point and by observing the ordinal pattern of the embedded time series from the original time series. It has been widely recognised as an efficient complexity measure due to its simplicity, fast calculation and invariance with respect to monotonous transformation. For details on the calculation and application of permutation entropy of time series, we refer to $[4]$. Among numerous complexity measures, extropy as discussed in [9] is also considered as a complexity measure which is related to entropy. Consider a random variable $X$ which takes values among $\{x_1,\ldots, x_n\}$ with respective probabilities  $\{p_{1},\ldots, p_{n}\}$, then the extropy associated with that random variable is defined as:

\begin{equation}\label{eqn 2}
J(X) = -\sum_{i=1}^{n} (\ 1- p_{i})\\\log(\ 1- p_{i} ).\
\end{equation}

Recently, Lad, Sanfilippo, and Agro [(2015), ‘Extropy: Complementary Dual of Entropy, Statistical Science, 30, 40–58] showed the measure extropy as a complementary dual of entropy. Here they briefly discussed about all the Shannon axioms and also compared between entropy and extropy.  Extropy of a probability mass function (pmf) is defined as the entropy of mass function which is complementary to that probability mass function. In this work, we suggest a different complexity measure, called permutation extropy and discuss its applications.

The paper is organized as follows. In Section $2$, the methodology has been discussed. Section $3$ deals with the simulation study. Here, we have considered logistic map, Henon map and Burger map. The finance and covid related data sets have been considered and analyzed in Sections $4$ and $5,$ respectively. Finally, Section $6$ concludes the paper. 

\section{Methodology}

Consider a time series $\{Y_{i}\}_{i=1}^{m}$, where $m$ denotes the length of the time series, taking values $\{y_1,\ldots,y_m\}$. Generally embedding is the relatively lower dimension into which we can translate the high dimensional vector. Based on this time series, we generate embedded time series representation $Y_{j}^{k}= \{y_j,\ldots,y_{j+k-1}\}$, for $j=1,\ldots, m-k+1$, where $k$ is the embedding dimension. Clearly by taking $k$ distinct numbers, we can generate maximum $k!$ possible ordinal patterns, and all patterns have unique ordering. Consider $\pi_{i}^{k}$ denotes the $i^{th}$ ordinal pattern generated from those $k$ distinct numbers, where ${i=1,\ldots,k!}$. Now, observe ordering of each time series out of those $m-k+1$ generated embedded time series will match with one of $k!$ possible ordinal patterns. Based on the extropy calculated in Equation $(2)$ permutation extropy is then defined as the extropy of $k!$ distinct symbols:
\begin{equation}
    J(k)=-\sum_{i=1}^{k!} (1-p(\pi_{i}^{k}))\{\log (1-p(\pi_{i}^{k}))\},
\end{equation}
where $p(\pi_{i}^{k})$ is defined as:
\begin{equation}
    p(\pi_{i}^{k})=\frac{\# \{j|j=1,\ldots,m-k+1:\mbox{type~}(Y_{j}^{k})=\pi_{i}^{k}\}}{m-k+1},
\end{equation}
where type(.) denotes mapping from a pattern space to the symbol space that is basically the ordinal pattern of the vector, $\#(.)$ denotes the cardinality of a set. Clearly the permutation extropy value will be in the range [0 , $(k!-1)(\log k!- \log (k-1)!)$]. The value will be equal to 0 when the time series is either completely increasing and decreasing and will equal to $(m!-1)(\log k!- \log (k-1)!)$, when there is equal likeness of happening of all $k!$ possible ordinal patterns.

\section{Simulation}
\subsection{Logistic Map}
Logistic map is a polynomial mapping of degree $2$, which is defined as:
\begin{equation}
    x_{n+1}=r x_{n}(1-x_{n}),
\end{equation}
where $x_{n}$ is any real number between 0 and 1 and parameter $r\in [0,4]$ which is also known as the control parameter. As known to all, logistic map has excellent properties in studying chaotic behavior. Fixing initial population, i.e. $x_0$, if we vary the parameter $r$ and generate the time series, we will eventually notice the value of $x$  attains some random values initially, after some iteration the equilibrium value $x$ will attain a fixed value or it will oscillate between some fixed values. With change of $r$, the data have different characteristics like periodic series, chaotic series, cluster points. Here, we choose $ r = 3, 3.1, 3.2, 3.3, 3.4, 3.5, 3.6, 3.7, 3.8 $ and the initial value $x_0=0.1$. 
\begin{figure}
\begin{subfigure}{0.5\textwidth}
\includegraphics[width=0.95\linewidth, height=7cm]{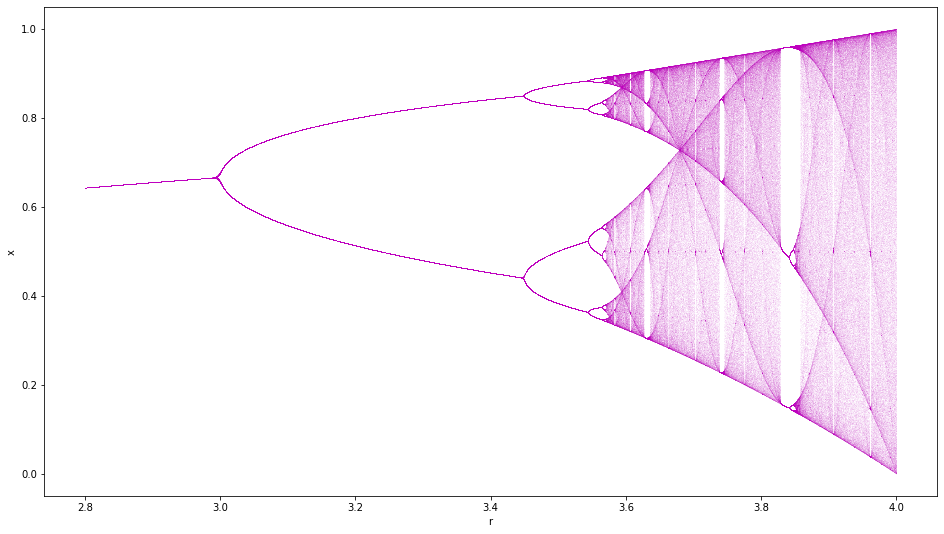} 
\caption{Bifurcation diagram for the logistic map.}
\label{fig:subim1}
\end{subfigure}
\begin{subfigure}{0.5\textwidth}
\includegraphics[width=0.95\linewidth, height=7cm]{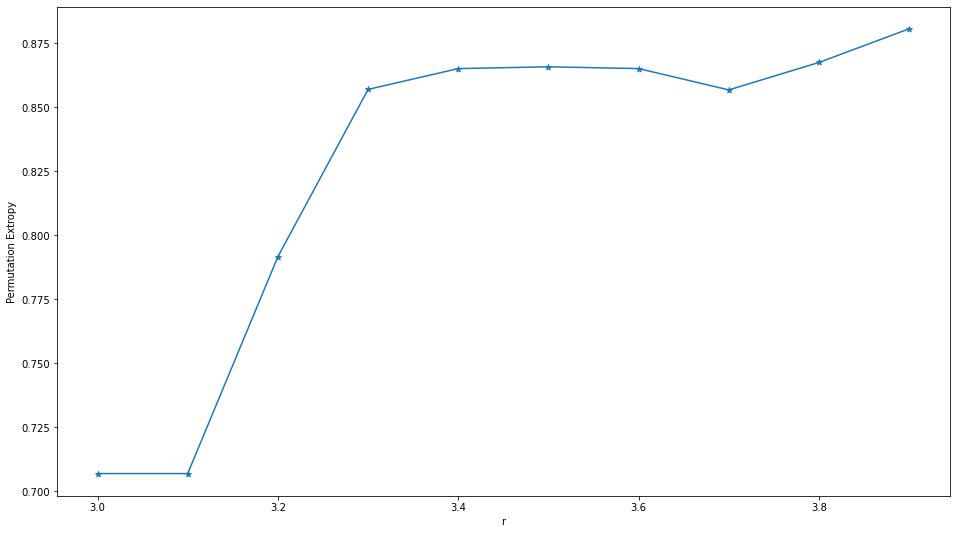} 
\caption{Permutation extropy of the logistic map.}
\label{fig:subim2}
\end{subfigure}
\caption{Bifurcation diagram and permutation extropy of logistic map.}
\label{fig:image0}
\end{figure}

First, we draw a bifurcation diagram (see Figure $1(a)$) of the logistic map by taking the initial value $x_0$ and varying different values of $r$ as discussed above. We observe from the bifurcation diagram that the data approach to a single value when $r=3$, approach to permanent oscillation between two values when $r= 3.2, 3.3, 3.4 $, and a stable cycle of period four when $r = 3.5$. Beyond $r = 3.5$, the data show periodic behavior at some values, and after $r = 3.8$ the chaotic behavior is observed. Now, we compute the permutation extropy values of different data series, which are generated by varying parameter $r$ and fixing the initial value $x_0$. After that, we plot the graph by taking parameter $r$ in $X$-axis and the permutation extropy values along $Y$-axis, which is shown in Figure 1(b). From this figure, we observe that when $r= 3, 3.1$, the permutation extropy values are nearly equal due to the fact that the data exhibit a same type of behavior. After that the permutation extropy value gradually increases with an increase in the parameter $r$ due to the observed increasing number of periods. Around $r=3.7$, the data exhibit periodic behavior although the data exhibit some kind of chaotic behavior earlier. Therefore, the permutation extropy value at $r=3.7$ decreases slightly. After that the permutation extropy value increases as usual. When the logistic map is fully chaotic, i.e. $r = 4$, the uncertainty or the degree of randomness is highest which implies the permutation extropy value will be the highest among all other $r$ values. When the logistic map exhibits chaotic behavior, the permutation extropy value is higher in comparison to the periodic behavior. Thus, We can state that the permutation extropy is a valid measure of uncertainty and has the potential for many applications. We plot permutation entropy of different time series generated by varying parameter $r$, and compare with permutation extropy values. The uncertainty should be more when $r=3.6$ in comparison to $r=3.7$, which is visible from the bifurcation diagram of the logistic map. As shown in Figure $2(a),$ the permutation entropy value at $r=3.7$ is higher than that at $r=3.6$. That implies permutation entropy measure can not precisely capture the complex behavior of a logistic map which is not the case for permutation extropy measure. Hence we can conclude that permutation extropy is an efficient measure than the permutation entropy. A comparison plot between the permutation entropy and permutation extropy has been depicted in Figure $2(b)$.

\begin{figure}
\begin{subfigure}{0.5\textwidth}
\includegraphics[width=0.95\linewidth, height=7cm]{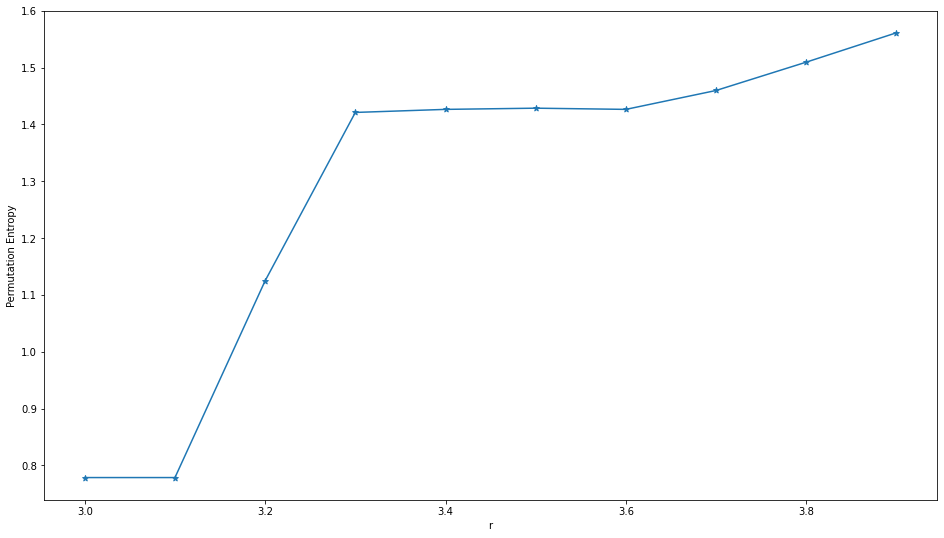} 
\caption{Permutation entropy of the logistic map.}
\label{fig:subim3}
\end{subfigure}
\begin{subfigure}{0.5\textwidth}
\includegraphics[width=0.95\linewidth, height=7cm]{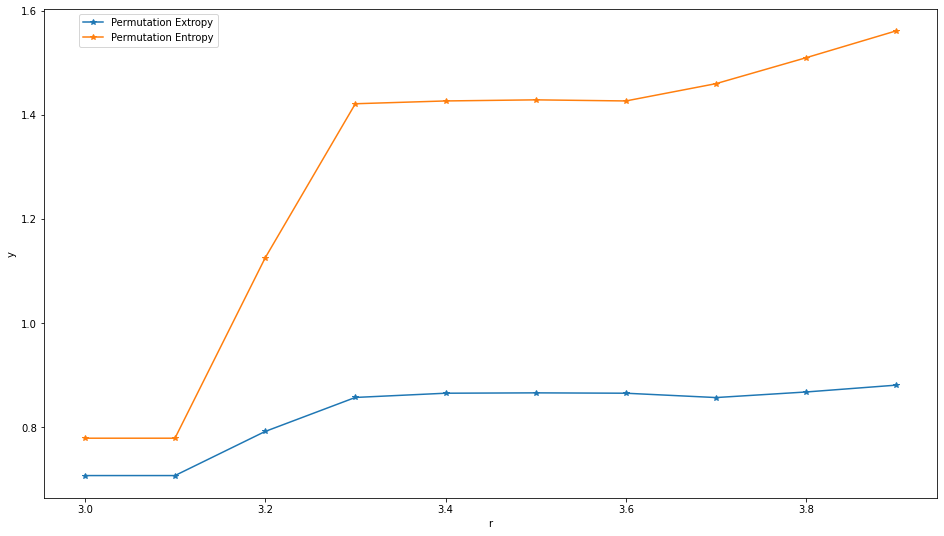} 
\caption{Complexity measures of the logistic map.}
\label{fig:subim4}
\end{subfigure}
\caption{Permutation entropy and comparison between permutation entropy and extropy of the logistic map.}
\label{fig:image1}
\end{figure}
\subsection{Henon Map}

Henon map, as discussed in [11] is a discrete time dynamical system which exhibits chaotic behavior and is defined as:
\begin{gather*} 
    x_{n+1}=1+y_{n}-a x_{n}^{2}\\
    y_{n+1}=b x_{n}.
\end{gather*}

\begin{figure}
\begin{subfigure}{0.5\textwidth}
\includegraphics[width=0.95\linewidth, height=7cm]{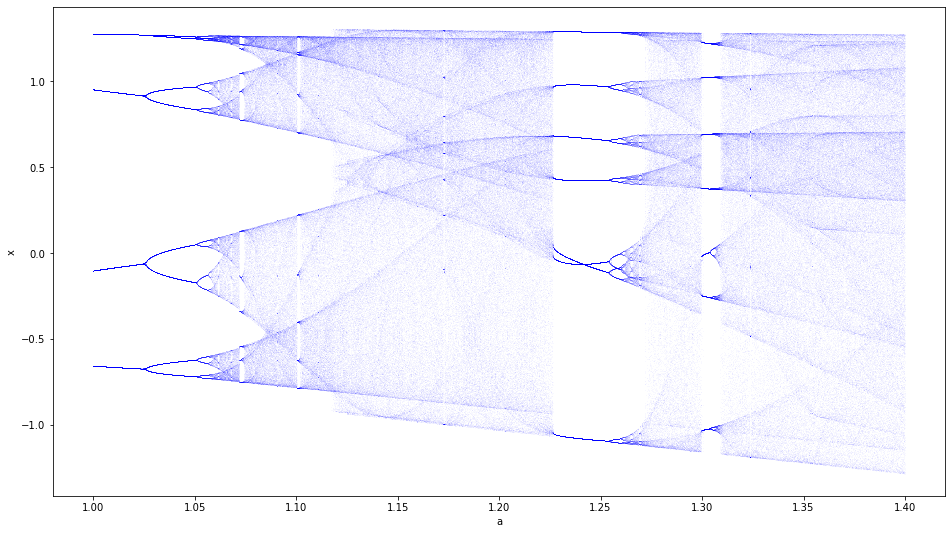} 
\caption{Bifuraction diagram for the Henon map when b=0.3.}
\label{fig:subim5}
\end{subfigure}
\begin{subfigure}{0.5\textwidth}
\includegraphics[width=0.95\linewidth, height=7cm]{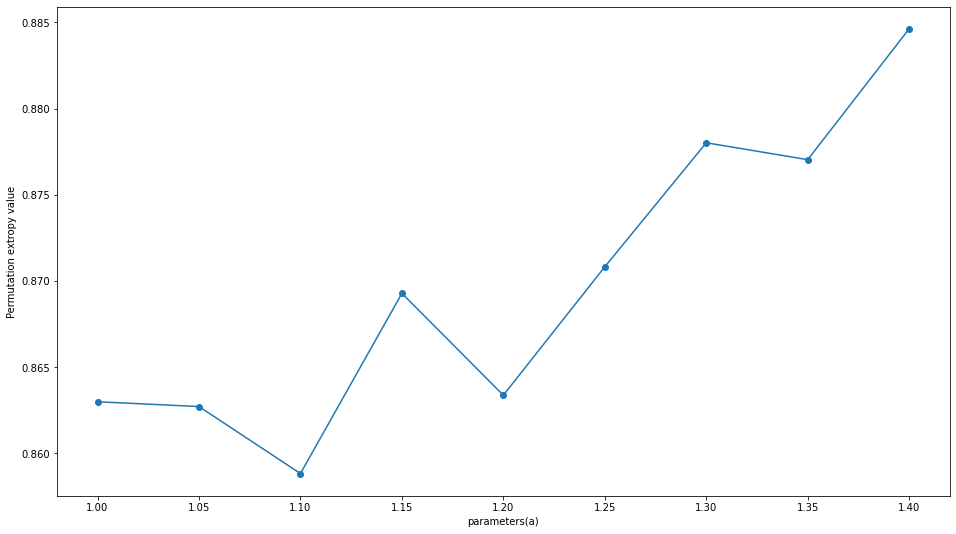} 
\caption{Permutation extropy of the Henon map when b= 0.3.}
\label{fig:subim6}
\end{subfigure}
\caption{Bifurcation diagram and permutation extropy of the Henon map.}
\label{fig:image3}
\end{figure}

The Henon map depends upon two parameters $a$ and $b$. By varying $a$ and $b$, we observe that the Henon map exhibits different behaviors/characteristics such as chaotic, periodic or intermittent. As known to all, the classical Henon map corresponds to $a =1.4$ and $b=0.3$ and it exhibits chaotic behavior. The Henon map can be deconstructed into  single dimension, which is given by
\begin{equation}
    x_{n+1}=1-a x_{n}^{2}+b x_{n-1}.
\end{equation}
First, by taking $b=0.3$, varying parameter $a$, i.e. $a \in [1,1.4] $ and fixing the initial condition $x_{0}=0.1, y_{0}=0.1$, we construct the Henon map. After that, we observe that the Henon map exhibits a mixing of periodic as well as chaotic behavior by increasing the parameter $a$. Now, by taking the value of parameter $b$ and the initial condition as discussed above, we construct different time series by varying $a$. After that, we calculate the permutation extropy of each time series generated by following the above procedure and construct a graph by taking parameter value along $X$ axis and permutation extropy values plotted along $Y$ axis. By looking at the graph, we can observe that by increasing the value of the parameter $a$, the permutation extropy value increases. When $a= 1, 1.05$, the permutation extropy values are nearly equal due to nearly the same degree of randomness. For some cases like, when $a= 1.1, 1.2, 1.35$ the trend in the graph changes due to the data series approach towards a periodic behavior other than chaotic behavior. When $a=1.4$, we notice more chaos, that is the uncertainty or degree of randomness is the highest among all the values of $a$. Thus, the permutation extropy value is more. As a result, the permutation extropy clearly implies/validates the characteristics of the Henon map. Further, its preciseness is more in comparison to other complexity measures.
\begin{figure}
\begin{subfigure}{0.5\textwidth}
\includegraphics[width=0.95\linewidth, height=7cm]{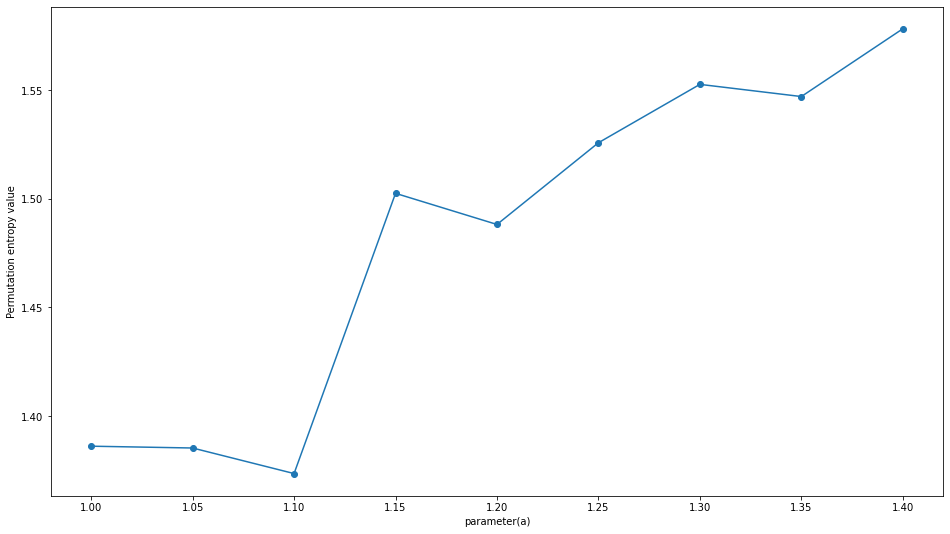} 
\caption{Permutation entropy for the Henon map when b=0.3.}
\label{fig:subim7}
\end{subfigure}
\begin{subfigure}{0.5\textwidth}
\includegraphics[width=0.95\linewidth, height=7cm]{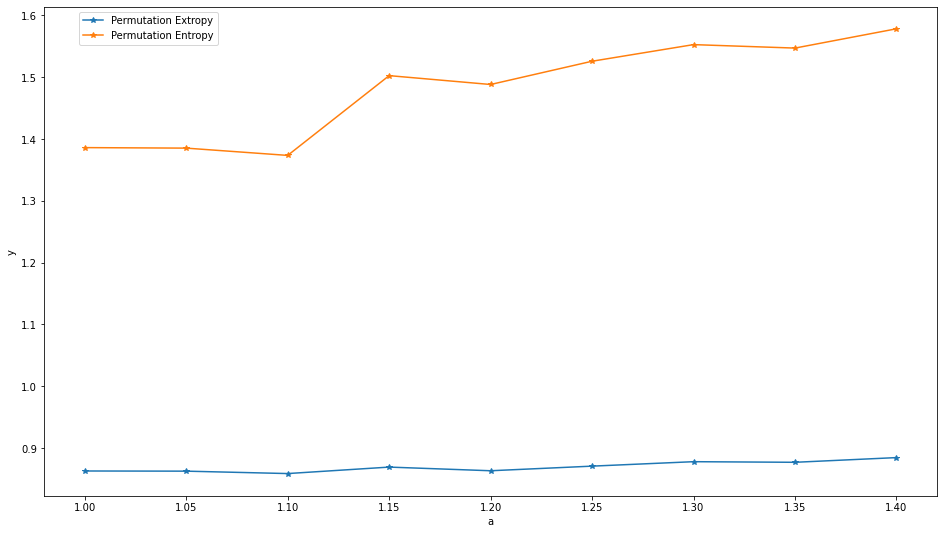} 
\caption{Permutation extropy and permutation entropy of the Henon map when b= 0.3.}
\label{fig:subim8}
\end{subfigure}
\caption{Permutation entropy and comparison between permutation entropy and extropy of the Henon map.}
\label{fig:image4}
\end{figure}

\begin{figure}
\begin{subfigure}{0.5\textwidth}
\includegraphics[width=0.95\linewidth, height=7cm]{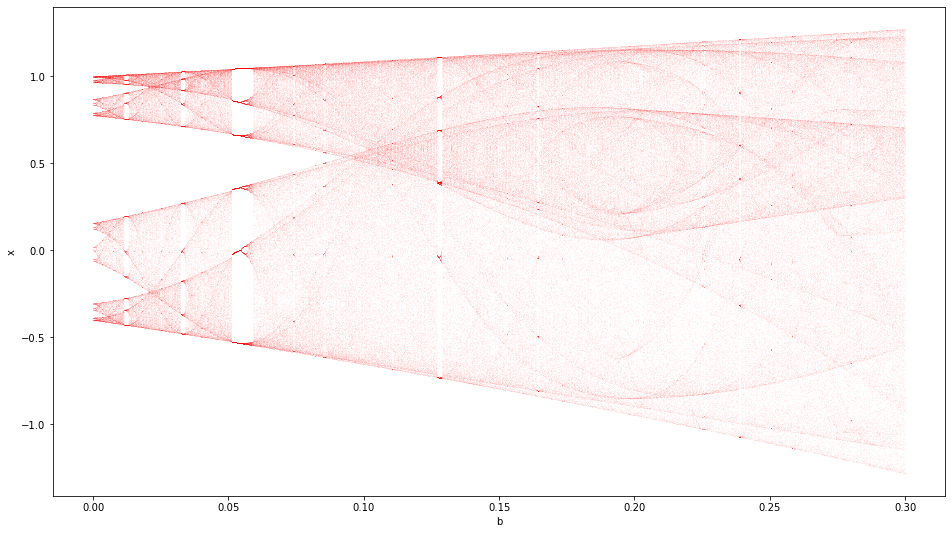} 
\caption{Bifurcation diagram for the Henon map when a=1.4.}
\label{fig:subim40}
\end{subfigure}
\begin{subfigure}{0.5\textwidth}
\includegraphics[width=0.95\linewidth, height=7cm]{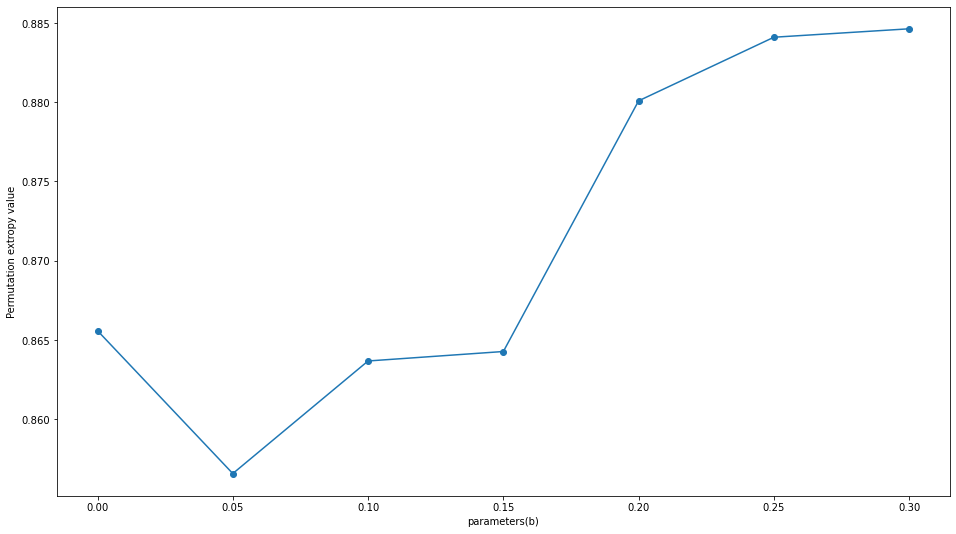} 
\caption{Permutation extropy of the Henon map when a=1.4.}
\label{fig:subim9}
\end{subfigure}
\caption{Bifurcation diagram and permutation extropy of the Henon map.}
\label{fig:image5}
\end{figure}

Similarly, we construct the bifurcation diagram of the Henon map (see Figure $3(a)$) by taking the same initial conditions and only varying parameter $b$ in place of parameter $a$ which takes a value of $1.4$. In the bifurcation diagram, we observe that the map exhibits both periodic and chaotic behavior. After plotting the graph in Figure $3(b)$ by taking $a$ values along $X$ axis and permutation extropy values along the $Y$ axis, we observe that the permutation extropy value gradually increases with the increase of $b$. When $b=0.05$, the map exhibits periodic behavior whose period value is less than the value when $b=0$. After that, the extropy value increases with an increase in $b$. When $b=0.3$, we see that chaos is more, that is the uncertainty or degree of randomness is the highest among all $a$. Therefore, the permutation extropy value is more. Thus, permutation extropy clearly implies/validates the characteristics of the Henon map and the preciseness is more in comparison to other complexity measures. We calculate the permutation entropy value of each time series generated by varying parameter $a$ and satisfying all the initial conditions as discussed above. After that, we plot the graph in Figure $4(a)$ by taking the permutation entropy value along $Y$ axis and parameter $a$ along $X$ axis. We compare two plots as shown together in Figure $4(b)$, hence we conclude that there is not much difference in the trend of the plots. As we observe the similarity in the case of fixing parameter $b$, now it is time for considering the case of fixing parameter $a$. Hence we compute the permutation entropy value of each time series generated by varying parameter $b$ and satisfying all the initial conditions as discussed above. After that, we plot the graph by taking the permutation entropy value along $Y$ axis and parameter $b$ along $X$ axis. While comparing both the figures as shown in Figure $5(b)$ and Figure $6(a)$, we observe some slightest change in the trend of the curve.  When $b=0.15$, the chaos is more in comparison to $b=0.10$ which can be clearly observed by looking at the bifurcation diagram of the Henon map as shown in Figure $5(a)$. That implies both permutation extropy and permutation entropy measures should be more in the case of $b=0.15$ in comparison to $b=0.10$. But by looking at Figure $6(a)$, it is clearly seen that the permutation entropy value is more when $b=0.10$, which is not in the case of permutation extropy. Thus, we conclude that permutation extropy more precisely captures the behavior of the Henon map in comparison to permutation entropy. In Figure $6(b)$, we plot both permutation extropy and permutation entropy, which is basically a comparison plot.

\begin{figure}
\begin{subfigure}{0.5\textwidth}
\includegraphics[width=0.95\linewidth, height=7cm]{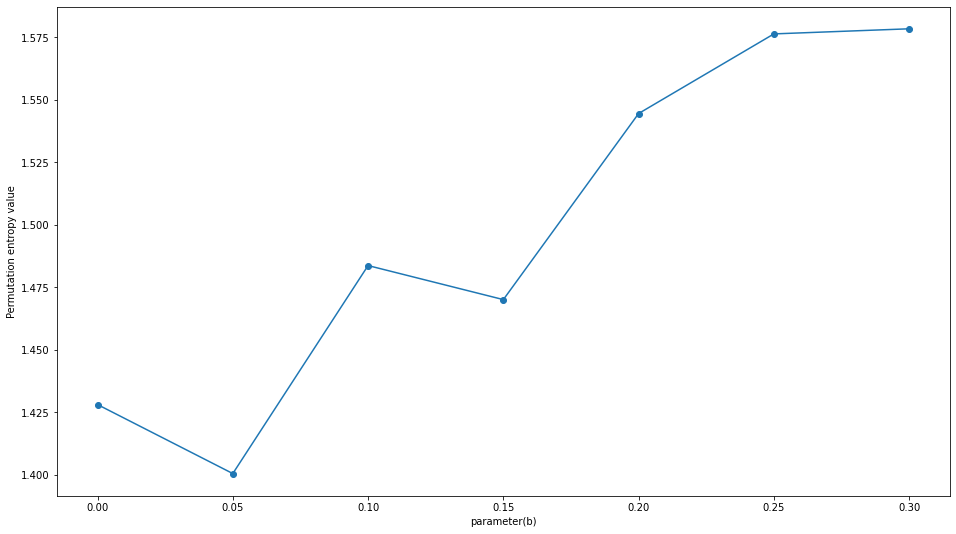} 
\caption{Permutation entropy for the Henon map when a=1.4.}
\label{fig:subim10}
\end{subfigure}
\begin{subfigure}{0.5\textwidth}
\includegraphics[width=0.95\linewidth, height=7cm]{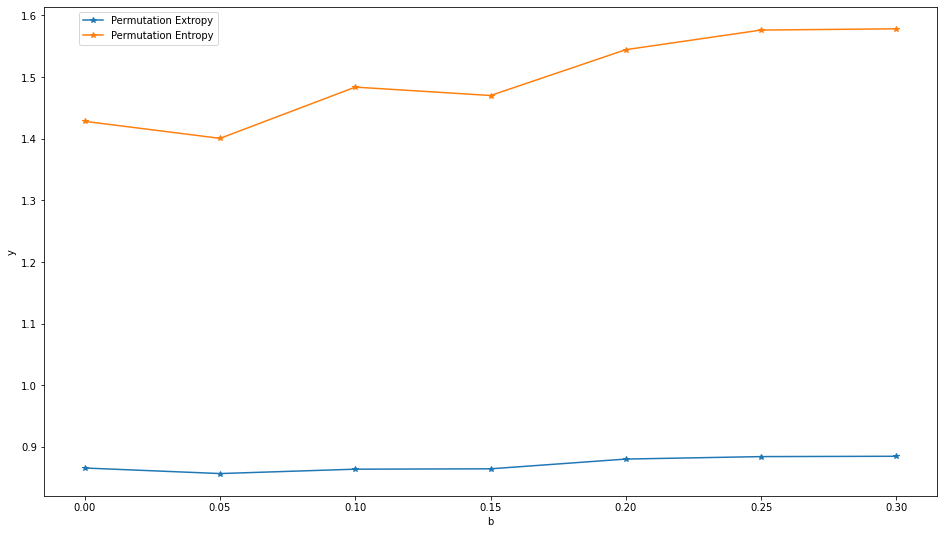} 
\caption{Permutation extropy and permutation entropy of the Henon map when a= 1.4.}
\label{fig:subim11}
\end{subfigure}
\caption{Permutation entropy and comparison between permutation entropy and extropy of Henon map.}
\label{fig:image6}
\end{figure}

\subsection{Burger Map}
Burger map is a discretization of a pair of coupled differential equations, which is defined as follows:
\begin{gather*} 
    x_{n+1}=(1-a)x_{n}-y_{n}^{2}\\
    y_{n+1}=(1+b)y_{n}+x_{n}y_{n},
\end{gather*}
where $a$ and $b$ are parameters. These equations are used to demonstrate the relevance of the bifurcation diagram of Burger map to the study of hydrodynamics fluid flow. E. M. ELabbasy, H. N. Agiza, H. EL-Metwally, A. A. Elsadany [10] studied the bifurcation analysis of Burger map in a detailed manner and displayed it's complex dynamics and chaotic behavior using a numerical study. In this subsection, we draw different bifurcation diagrams by varying parameters $a$ and $b$. After that we use the concept of permutation extropy measure to validate chaotic and periodic behaviors of the map, and then compare the extropy measure with permutation entropy.
\begin{figure}
\begin{subfigure}{0.5\textwidth}
\includegraphics[width=0.95\linewidth, height=7cm]{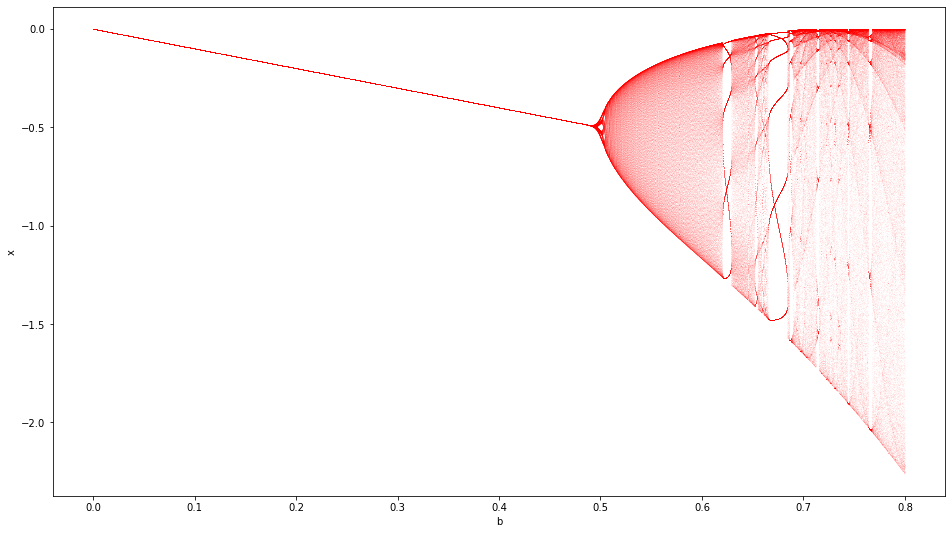} 
\caption{Bifurcation diagram for the Burger map when a=1.}
\label{fig:subim12}
\end{subfigure}
\begin{subfigure}{0.5\textwidth}
\includegraphics[width=0.95\linewidth, height=7cm]{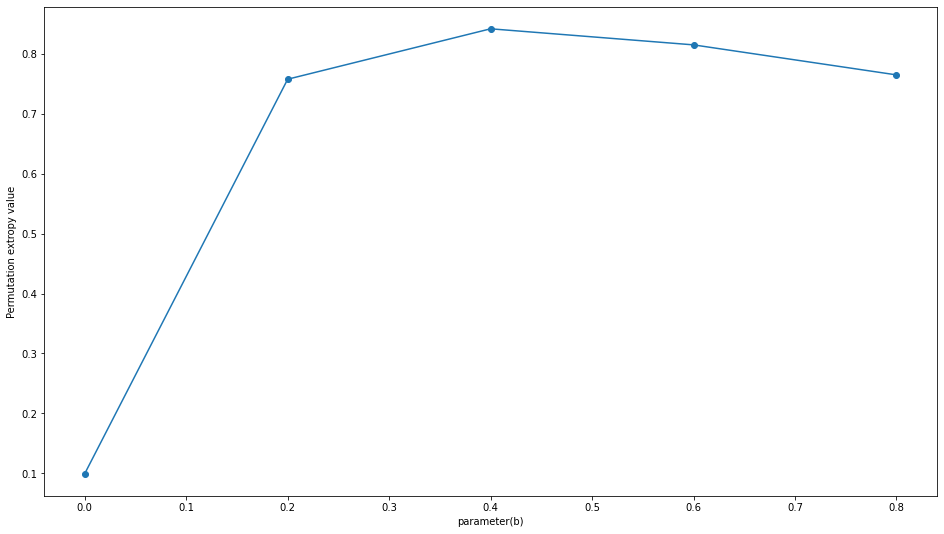} 
\caption{Permutation extropy of the Burger map when a=1.}
\label{fig:subim13}
\end{subfigure}
\caption{Bifurcation diagram and permutation extropy of the Burger map.}
\label{fig:image7}
\end{figure}

\begin{figure}
\begin{subfigure}{0.5\textwidth}
\includegraphics[width=0.95\linewidth, height=7cm]{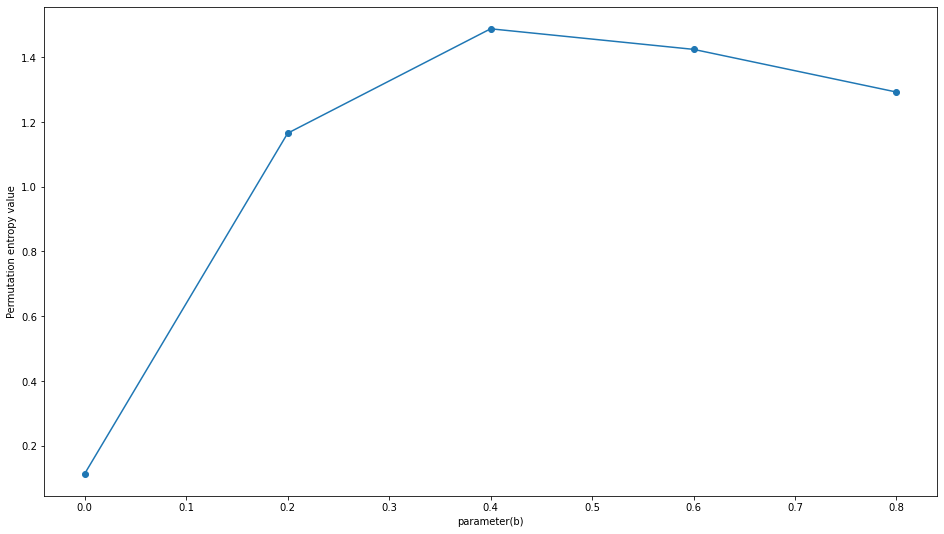} 
\caption{Permutation entropy for the Burger map when a=1.}
\label{fig:subim14}
\end{subfigure}
\begin{subfigure}{0.5\textwidth}
\includegraphics[width=0.95\linewidth, height=7cm]{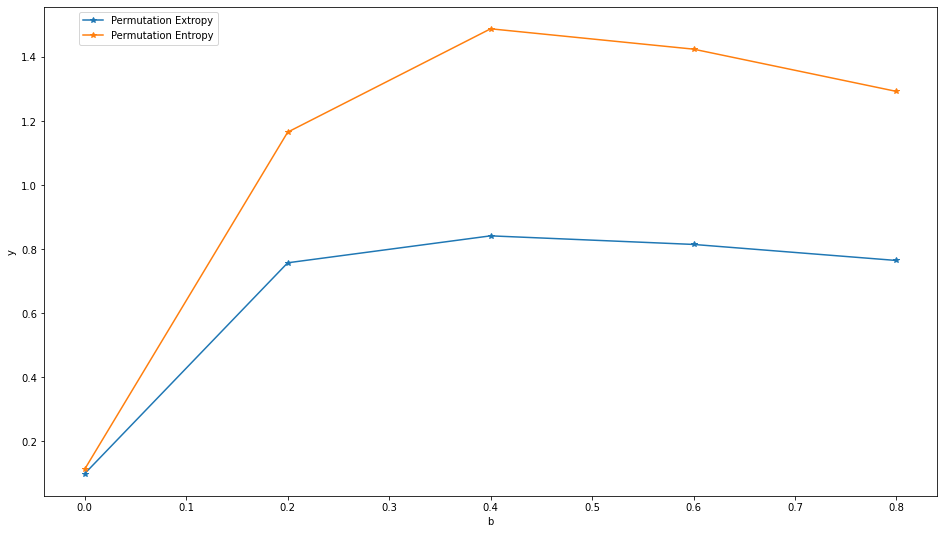} 
\caption{ Comparison diagram.}
\label{fig:subim15}
\end{subfigure}
\caption{Permutation entropy and comparison between permutation entropy and extropy of the Burger map.}
\label{fig:image8}
\end{figure}
We plot the bifurcation diagram of the burger map (see Figure $7(a)$) by taking parameter $a=1$, initial values $x_{0}=1, y_{0}=1$ and varying the parameter $b$ in $[0,0.8]$. We generate different time series corresponding to different parameters $b$ by assuming all the conditions that we take during the plot of the bifurcation diagram. After that, we calculate the permutation extropy of different time series by following the steps for calculating permutation extropy as described in Section $2$. We plot the graph by taking parameter $b$ along $X$ axis and permutation extropy value along $Y$ axis. While computing the permutation extropy of a time series, we generate sub time series by fixing a particular embedding dimension value, suppose the value is $m$. We get different permutation patterns using different $m$ values and evaluate the probability of each permutation pattern, which is briefly described in Section $2$. Suppose for a particular time series, we get a probability value of $1$ for a particular permutation pattern and a probability value of $0$ for all other remaining patterns, which can clearly imply that there is no chaos and hence the permutation value should be low. We will also get a low permutation extropy value in the time series where there is less probability of occurring each permutation pattern, in this case, we observe periodic behavior or in case if we observe some chaos, then the chaos will be less. When the probability of occurrence of each generated permutation pattern is high in comparison to the probability of each permutation pattern from other time series, the map will exhibit highly chaotic behavior and hence the permutation extropy should be high in comparison to others. While taking $b=0$, we observe that although  we will get some positive probability of occurrence of some permutation patterns, the probability of occurrence of some permutation patterns is equal to $0$. Hence there is less chaos, i.e. permutation extropy value should be less which is reflected in Figure $7(b)$. While taking parameter $b=0.4$, we observe that the probability of each permutation pattern is high in comparison to the remaining permutation patterns, which is clearly visible in Figure $7(b).$ Following the arguments stated above, we obtain the trend in the curve as shown in Figure $7(b).$ Similarly, we calculate the permutation entropy of each time series generated by varying parameter $b$ and assuming the initial condition. While plotting the graph by taking permutation entropy along $Y$ axis and parameter $b$ along $X$ axis, we observe the slightest change in the trend of the graph. As per the argument we discussed above, the permutation extropy value at $b=0.4$ should be not that much larger than the value at $b=0.2$, which is clearly visible in Figure $7(b)$. In Figure $8(a)$, the permutation entropy could not capture that preciseness. But when $b=0.6$ and $b=0.8$, both graphs exhibit the same kind of behavior. Here, we can say that although the permutation entropy is a complexity measure, the measure permutation extropy is a little bit more reliable. We draw the bifurcation diagram by taking all the initial conditions and parameter values while plotting the bifurcation diagram in Figure $7(a),$ but here the only difference is that we consider the variation of $y$ series that we get by using the definition of Burger map. After that, we calculate the permutation extropy of different time series by following the steps for calculating permutation extropy as described in Section $2$. We plot the graph by taking parameter $b$ along $X$ axis and permutation extropy value along $Y$ axis. When $b=0$, the probability value of two permutation patterns is positive and all other patterns have $0$ probability. Hence the chaos is less, and we can say there is nearly no chaos which implies the permutation extropy value must be low in comparison to others because we get a positive probability of occurrence of each permutation pattern while taking different values of $b$. When $b=0.4$, the probability of occurrence of each permutation pattern is high in comparison to $b=0.2$, which implies the permutation extropy value at $b=0.4$ in comparison to $b=0.2$. When $b=0.6$ and $b=0.8$, the probability of occurrence of each permutation pattern is nearly equal, which implies the permutation extropy value at both $b$ values should not differ that much. In Figure $9(a)$, the permutation extropy values at $b=0.6$ and $b=0.8$ are nearly equal. The extropy value at $b=0.8$ is slightly less than $b=0.6$. Following the argument stated above, we get the trend in the curve as shown in Figure $7(b)$. Similarly, we calculate the permutation entropy of each time series generated by varying parameter $b$ and assuming the initial condition. While plotting the graph by taking permutation entropy along $Y$ axis and parameter $b$ along $X$ axis, we observe the slight change in the trend of the graph. We observe that the probability of occurrence of all permutation patterns is equal expect one pattern whose probability value is significantly changed i.e. high when $b=0.8$ in comparison to $b=0.6$. So the value of the complexity measure should be slightly higher in the case of $b=0.8$ in comparison to $b=0.6$, which is clearly reflected in Figure $10(a)$ but not in the case of Figure $9(b)$. When $b=0.4$, the probability of occurrence of each permutation pattern is high in comparison to $b=0.2$, which implies the permutation extropy value at $b=0.4$ in comparison to $b=0.2$. But the complexity measure value at $b=0.4$ should not be that much high in comparison to $b=0.4$. In Figure $9(b)$, the permutation extropy measure follow the exact argument as discussed above but which is not in the case of permutation entropy.\\
 We plot the bifurcation diagram of the Burger map by fixing parameter $b=0$, the varying parameter $a$ and assuming all the initial conditions that we take at the time of plotting Figure $7(a)$. When we vary the parameter $a$ in the range $[0,2]$, we get the bifurcation diagram as shown in Figure $11(a)$.
 If we observe Figure $11(a)$, we can not get the exact behavior, i.e. chaotic and periodic of the Burger map. If we vary the parameter $a$ in the range $[1,1.4]$, we can clearly observe some interesting patterns that are exhibited by the Burger map. We evaluate the permutation extropy of different time series that we get by the varying parameter $a$ in the range $[1,1.4]$.

\begin{figure}
\begin{subfigure}{0.5\textwidth}
\includegraphics[width=0.95\linewidth, height=7cm]{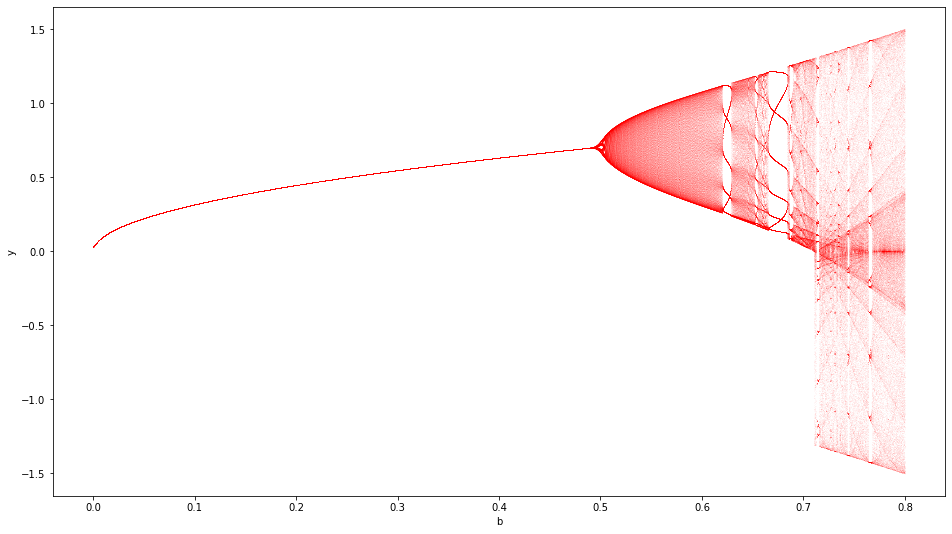} 
\caption{Bifurcation diagram for the Burger map when a=1.}
\label{fig:subim16}
\end{subfigure}
\begin{subfigure}{0.5\textwidth}
\includegraphics[width=0.95\linewidth, height=7cm]{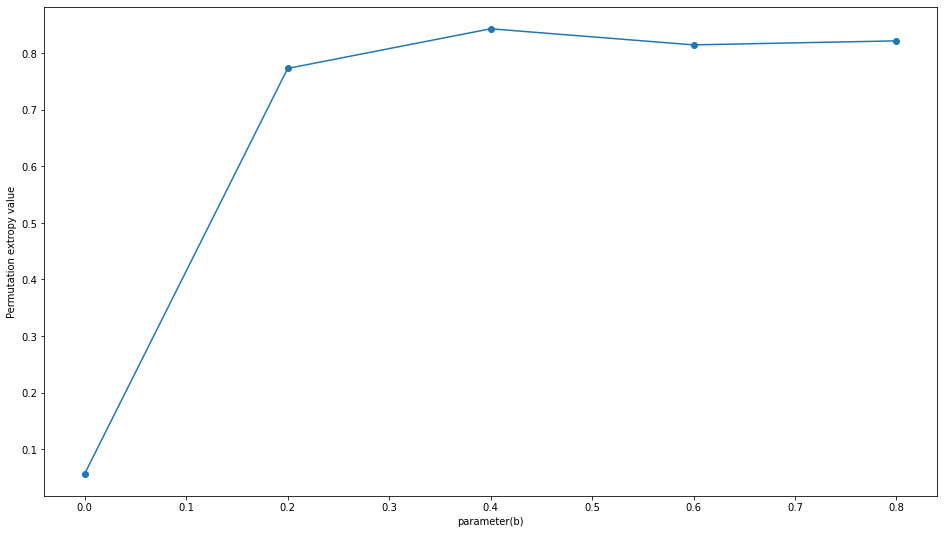} 
\caption{Permutation extropy of the Burger map when a=1.}
\label{fig:subim17}
\end{subfigure}
\caption{Bifurcation diagram and permutation extropy of the Burger map.}
\label{fig:image28}
\end{figure}

\begin{figure}
\begin{subfigure}{0.5\textwidth}
\includegraphics[width=0.95\linewidth, height=7cm]{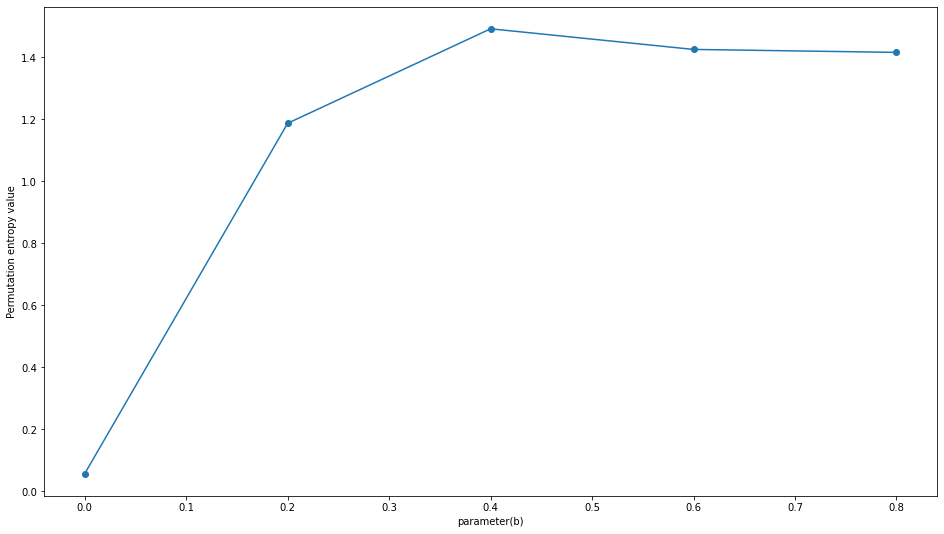} 
\caption{Permutation entropy for the Burger map when a=1.}
\label{fig:subim18}
\end{subfigure}
\begin{subfigure}{0.5\textwidth}
\includegraphics[width=0.95\linewidth, height=7cm]{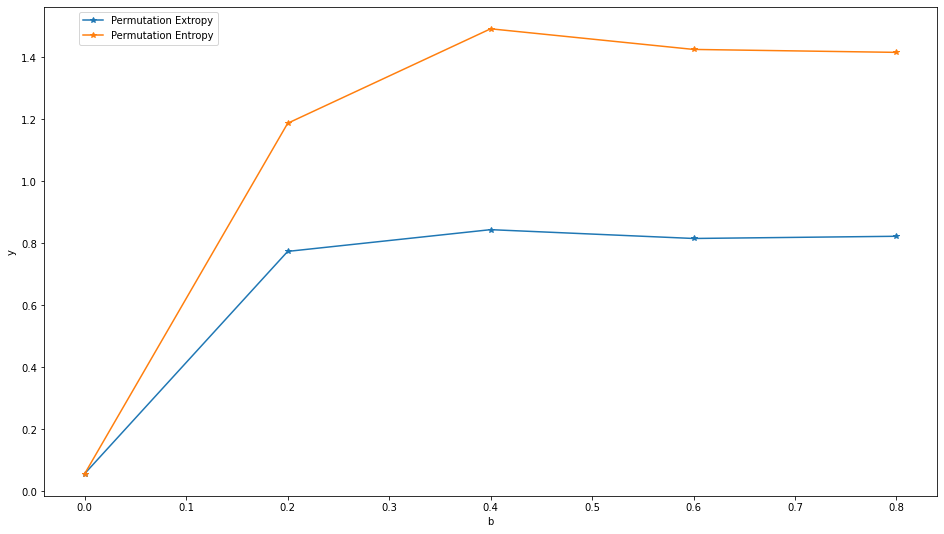} 
\caption{ Comparison diagram.}
\label{fig:subim19}
\end{subfigure}
\caption{Permutation entropy and comparison between permutation entropy and extropy of the Burger map.}
\label{fig:image9}
\end{figure}

\begin{figure}
\begin{subfigure}{0.5\textwidth}
\includegraphics[width=0.95\linewidth, height=7cm]{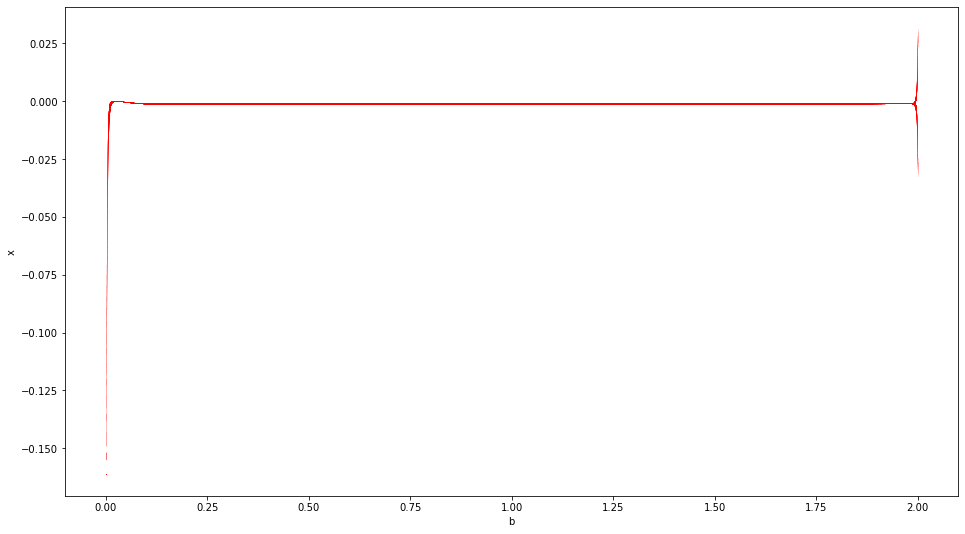} 
\caption{Bifurcation diagram for the Burger map when b=0.}
\label{fig:subim20}
\end{subfigure}
\begin{subfigure}{0.5\textwidth}
\includegraphics[width=0.95\linewidth, height=7cm]{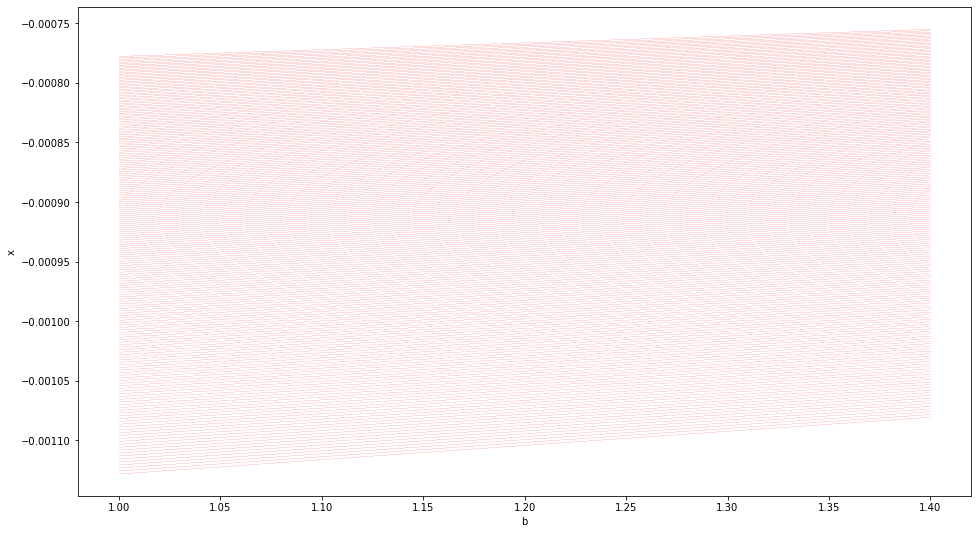} 
\caption{Bifurcation diagram for the Burger map when b=0.}
\label{fig:subim21}
\end{subfigure}
\caption{Bifurcation diagram of the Burger map.}
\label{fig:image10}
\end{figure}

\begin{figure}
\begin{subfigure}{0.5\textwidth}
\includegraphics[width=0.95\linewidth, height=7cm]{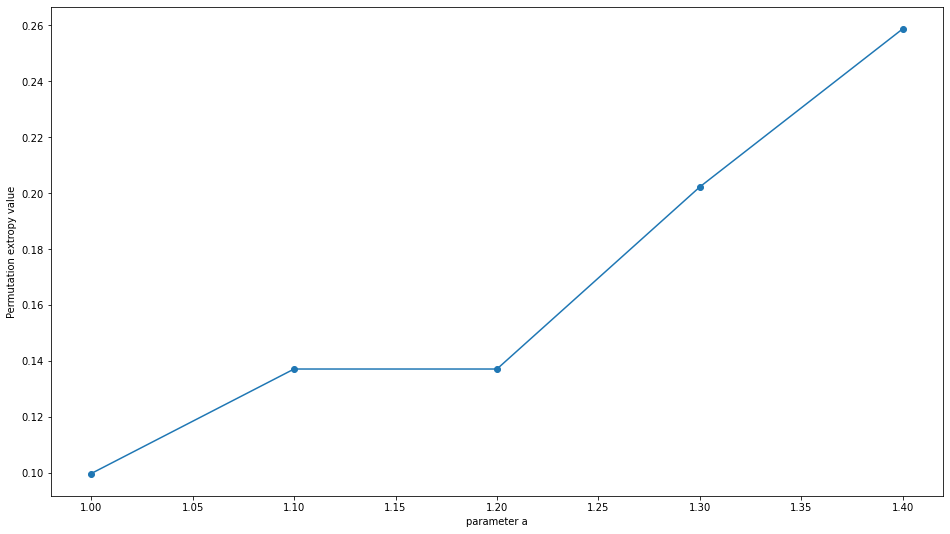} 
\caption{Permutation entropy for the Burger map when b=0.}
\label{fig:subim22}
\end{subfigure}
\begin{subfigure}{0.5\textwidth}
\includegraphics[width=0.95\linewidth, height=7cm]{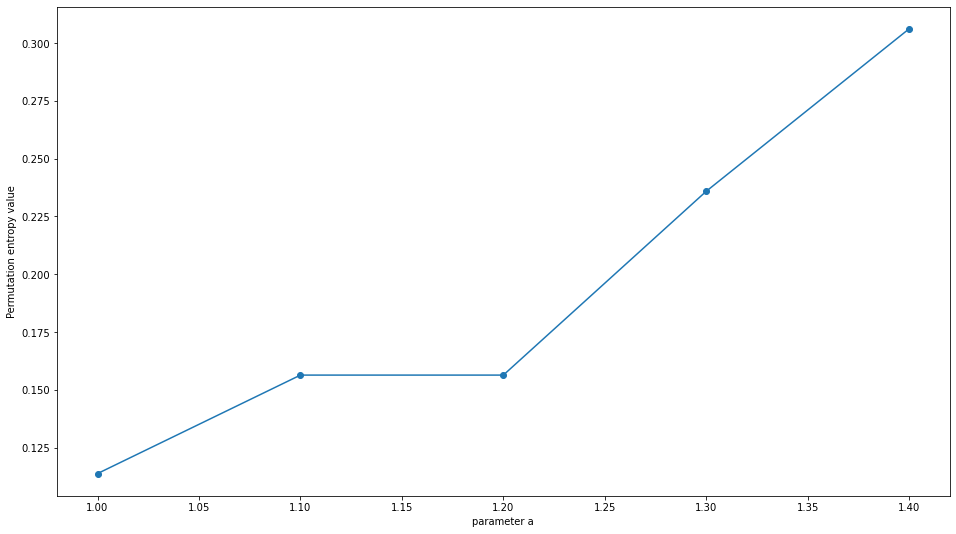} 
\caption{ Permutation entropy for the Burger map.}
\label{fig:subim23}
\end{subfigure}
\caption{Permutation entropy and permutation extropy of the Burger map.}
\label{fig:image11}
\end{figure}
We plot the graph by taking parameter $a$ along $X$ axis and permutation extropy value along $Y$ axis. If we observe the bifurcation diagram as shown in Figure $11(b)$ when $a=1$ there is less chaos in comparison to other $a$ values. We can state that there is less chaos when $a=1$, because of the patterns we observed in Figure $11(b)$. When $a=1.1$ and $a=1.2$, we observe nearly similar patterns. Hence chaos is nearly the same and the permutation extropy values are also equal. When $a=1.4$, the chaos is more than we can clearly observe from Figure $11(b)$. Hence the permutation extropy is the highest in comparison to other $a$ values. Similarly, we plot the graph by taking permutation entropy along $X$ axis and taking parameter $a$ along $Y$ axis, which is shown in Figure $12(b)$. We observe a similar kind of trend as we observe in the permutation extropy graph. We plot the bifurcation diagram of the burger map by fixing parameter $a=0.5$ and varying parameter $b$ as shown in Figure $13(a)$. Similarly we plot the bifurcation diagram by taking parameter $a=2$. For different values of parameter $a$, we plot the graph of permutation extropy which is shown in Figure $14$.
\begin{figure}

\begin{subfigure}{0.5\textwidth}
\includegraphics[width=0.95\linewidth, height=7cm]{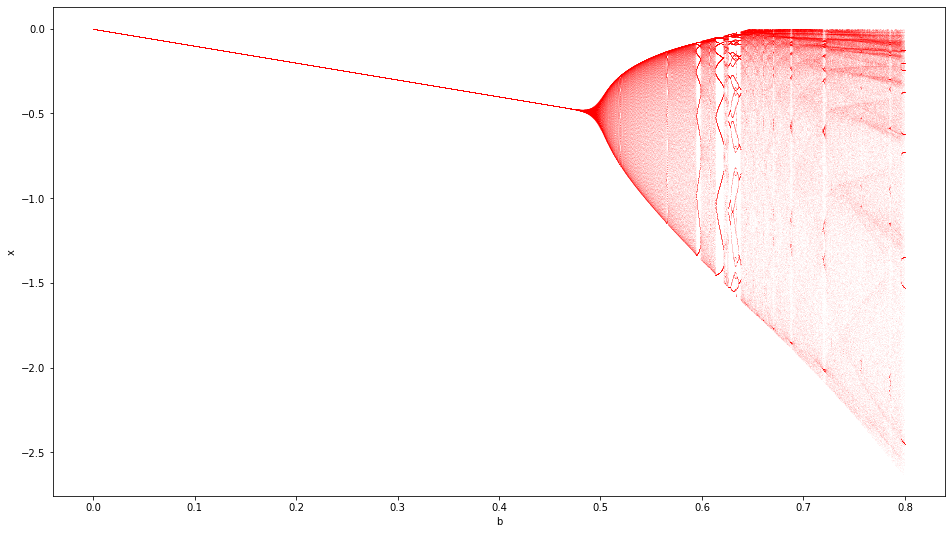} 
\caption{Bifurcation diagram  when a=0.5.}
\label{fig:subim24}
\end{subfigure}
\begin{subfigure}{0.5\textwidth}
\includegraphics[width=0.95\linewidth, height=7cm]{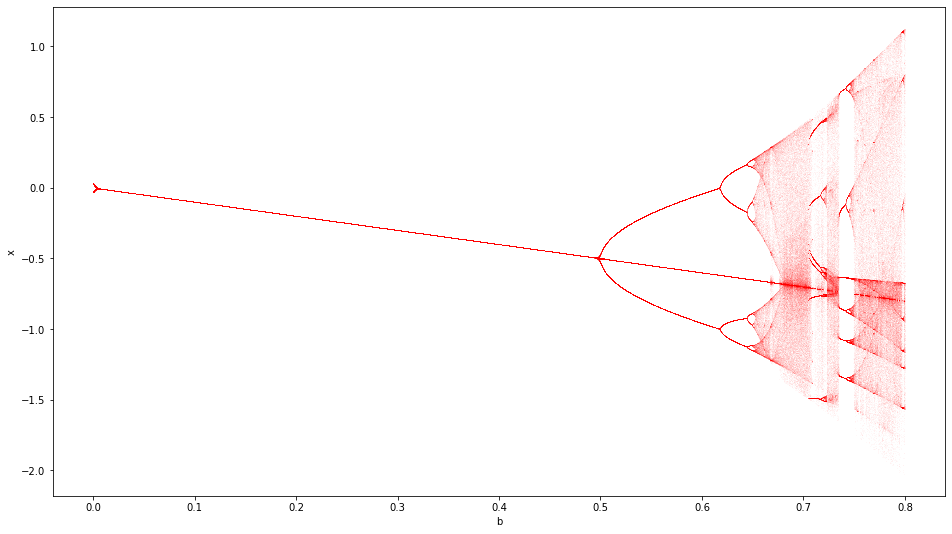} 
\caption{Bifurcation diagram for the Burger map when a=2.}
\label{fig:subim25}
\end{subfigure}
\caption{Bifurcation diagram of the Burger map.}
\label{fig:image12}
\end{figure}

\begin{figure}

\includegraphics[width=0.95\linewidth, height=7cm]{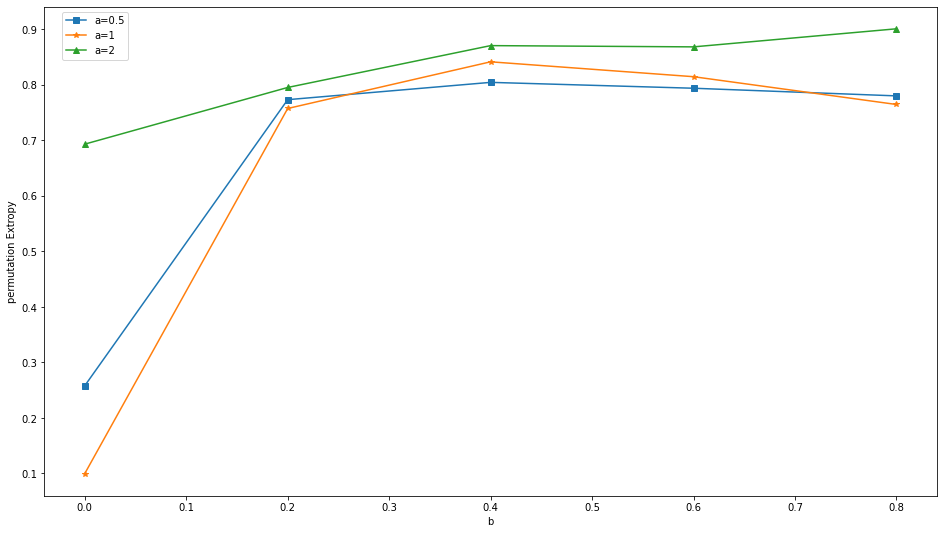} 
\caption{Comparison of permutation extropy  taking different values of $a$.}
\label{fig:subim26}

\end{figure}

\section{Financial Data Analysis}
Here we analyze the impact of covid on financial stock market with the help of the complexity measure permutation extropy. We take the data of closing price [12] of NIFTY50 that is the price of each day over four years i.e. 2019, 2020, 2021, 2022. Basically NIFTY 50 is an index of Indian stock market which represents the weighted average of 50 of the largest Indian companies listed on the National Stock Exchange. We discuss about the closing price and it's importance. During a particular day, the price in which the stock market closes for that day is called closing price for that particular day. With the help of closing prices we can compare the value of stock of one day with the previous trading day, which is very helpful for the investors. Closing price is also considered as the most accurate measure of the price of particular stock on the considered trading day. By comparing the closing price of a particular day with the previous day or suppose 30 days or may be a year, we can get an overall idea about the market sentiment towards  stock market. Here market sentiment refers to the behavior of the investors towards a particular stock.  First of all we take the data across four different financial years to analyze the behavior of the stock market or the market sentiment. We take the closing price data of each day of year 2018. First of all we generate the time series using the closing price data. After that using the methodology described above for calculating the permutation extropy, we calculate the permutation extropy of closing price series for the year 2018. Similarly, following these steps, we obtain the permutation extropy of different financial years i.e. 2019, 2020, 2021, 2022. Then we try to analyze by taking all the permutation extropy values and plotting these values in the graph which is shown in Figure $16(a)$. We clearly distinguish the behavior of the market between the years when covid is not there and the years when covid is at peak. Clearly during the year 2018 and 2019, the uncertainty in the stock market is more that we can observe using closing price data, hence the complexity measure i.e. the permutation extropy is more. The uncertainty in the year 2019 is more compared to 2018, thus the permutation extropy value is slightly more in comparison to 2018. In the year 2020, the covid started and during this period if we observe the behavior of the closing price, then the price fluctuate around the same value where the amount of fluctuation is very less. For each day, the stock market close at nearly same price and the uncertainty of the stock market is very less, that implies the permutation extropy value should be less. That we can clearly see, that is the permutation extropy value in the year 2020 is less in comparison to other financial years in Figure $16(b)$. In the year 2021, the impact of covid is slightly less than 2020. Hence the complexity is more because of the behavior of the stock market. The permutation extropy value is more in the year 2021 in comparison to 2020. After that we try to analyze the behavior of the stock market in the year 2022. Although covid is not there, but there is impact of it on the stock market. The permutation extropy value in the year 2022 is nearly equal to that of the year 2021 and slightly more than the year 2020, which can be clearly seen from Figure $15.$
\begin{figure}
\begin{subfigure}{0.5\textwidth}
\includegraphics[width=2.05\linewidth, height=7cm]{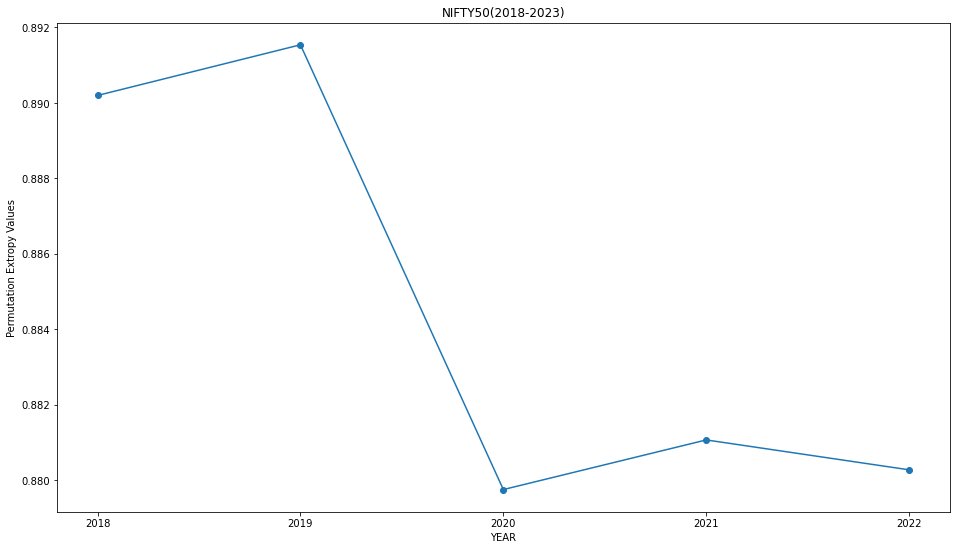} 
\label{fig:subim27}
\end{subfigure}
\caption{Permutation extropy plot of different financial years.}
\label{fig:image13}
\end{figure}

\begin{figure}

\begin{subfigure}{0.5\textwidth}
\includegraphics[width=0.95\linewidth, height=7cm]{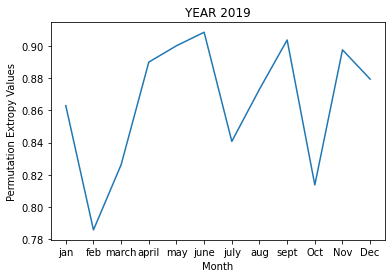} 
\caption{Financial Year 2019.}
\label{fig:subim28}
\end{subfigure}
\begin{subfigure}{0.5\textwidth}
\includegraphics[width=0.95\linewidth, height=7cm]{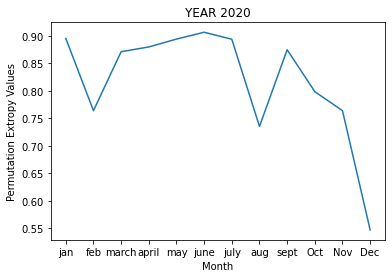} 
\caption{Financial Year 2020.}
\label{fig:subim29}
\end{subfigure}
\begin{subfigure}{0.5\textwidth}
\includegraphics[width=0.95\linewidth, height=7cm]{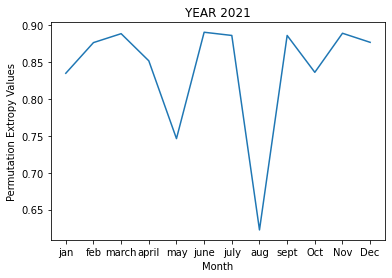} 
\caption{Financial Year 2021.}
\label{fig:subim30}
\end{subfigure}
\begin{subfigure}{0.5\textwidth}
\includegraphics[width=0.95\linewidth, height=7cm]{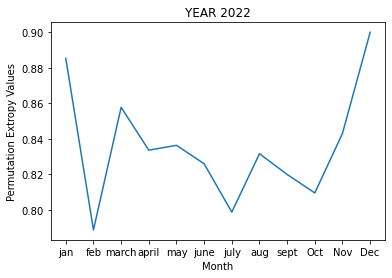} 
\caption{Financial Year 2022.}
\label{fig:subim31}
\end{subfigure}
\caption{Permutation extropy value of different months during a financial year (2019-2022).}
\label{fig:image14}
\end{figure}
We try to analyze the behavior of stock market in details manner by taking the closing price data and calculating the permutation extropy value of each month during a financial year. For each month, we take the closing price data for each day and treat the data set as a time series. After that using the methodology defined above in the Section 2 we calculate the permutation extropy of that month. We plot the graph by taking different months during a year along $X$ axis and permutation extropy values along $Y$ axis, which is clearly shown in the Figure 16. In Figure 16(a), the permutation extropy value in January is more than February, after that the extropy value increases up to May, after that the extropy value decreases up to the month July, the value in August is more in comparison to July, then it decreases afterwards up to October and then increases. The extropy values of November and December are nearly equal. During the year 2019, the permutation extropy constantly fluctuates between different values and the amount of fluctuation is more. That implies the uncertainty is more in the year 2019. During the year 2020, the permutation extropy value of the month February is less than January. From March to July, the permutation extropy values are nearly equal. The value of August is less in comparison to July. After that the value increases up to September and after September the value decreases up to December. In the year 2020, there is no much fluctuation in permutation extropy values. Hence the uncertainty in 2020 is less in comparison to the year 2019. Now for the year 2021, the permutation value increases up to March and afterwards it decreases up to May. After May, the value increases and the extropy value remains constant for two months, i.e. June and July. After July, the extropy value certainly decreases up to a large extent and after that the extropy values are nearly equal except some slight deviation in the value in the month October. For the financial year 2021, the permutation extropy values of all months are nearly equal except some slight fluctuation and except the month May and August. Hence, there is no much uncertainty in the year 2021. In the year 2022, the permutation extropy value of the month January is nearly equal to that of the month December of 2021. After January, the extropy value decreases up to February and the extropy value increases afterwards.  From March to  October, the extropy values do not vary too much except the months April and July. After October, the extropy value increases in large extent  up to December. Although, the fluctuation in the year 2022 is not that much but is more in comparison to the year 2021. So, the uncertainty in the stock market in the year 2022 is more in comparison to the year 2021. From this detailed analysis of the behavior by calculating permutation extropy of each month of a specific year, we also observe that the uncertainty in the market is less during the year 2020 and 2021, where the covid is at peak in comparison to the year 2019 and 2022. Hence the permutation extropy is a valid measure for giving remarks about the behaviour of the stock market.

\section{WHO Data Analysis}

\begin{figure}
\begin{subfigure}{0.5\textwidth}
\includegraphics[width=0.95\linewidth, height=7cm]{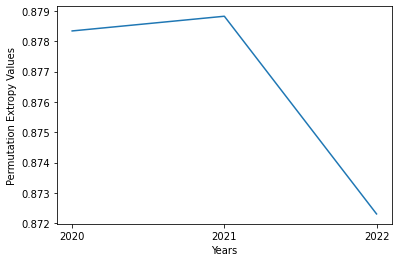}
\caption{Year 2020-2022.}
\label{fig:subim32}
\end{subfigure}
\begin{subfigure}{0.5\textwidth}
\includegraphics[width=0.95\linewidth, height=7cm]{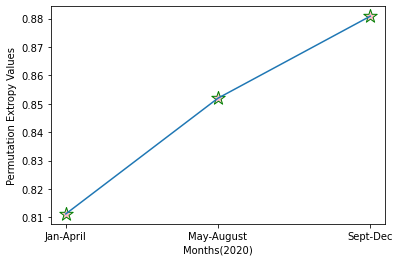} 
\caption{ Year 2020.}
\label{fig:subim33}
\end{subfigure}
\begin{subfigure}{0.5\textwidth}
\includegraphics[width=0.95\linewidth, height=7cm]{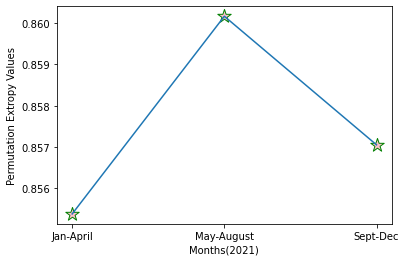} 
\caption{ Year 2021.}
\label{fig:subim34}
\end{subfigure}
\begin{subfigure}{0.5\textwidth}
\includegraphics[width=0.95\linewidth, height=7cm,right]{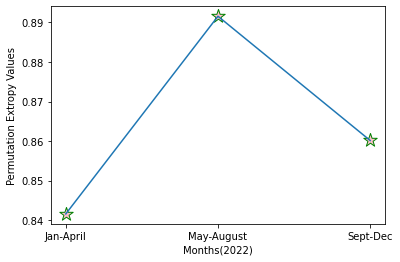} 
\caption{ Year 2022.}
\label{fig:subim35}
\end{subfigure}
\caption{Permutation extropy value of different years and  phases during a  year (2020-2022).}
\label{fig:image15}
\end{figure}
In this section, we study the situation of covid in India across different phases or years by taking the data [13] from WHO. We take the data of the number of deaths in a day across a year and treat it as time series. After that using the methodology described in Section 2 we calculate the permutation extropy of each year. We plot the graph by taking the year along $X$ axis and the permutation extropy value along $Y$ axis, which is shown in Figure $17(a).$ When we calculate the permutation extropy for the year 2020 by using WHO data, the value is less than that of the year 2021 or we can say nearly equal value. There is slightly more fluctuation in the number of death cases in 2021 compared to the year 2020. But when we calculate the permutation extropy value for the year 2022, it is very less in comparison to other years. The reason may be during the year 2022, the impact of covid is less due to several measures taken like vaccination and many more. Therefore, the number of death cases significantly decrease in comparison to the years 2020 and 2021.
\begin{figure}

\includegraphics[width=0.95\linewidth, height=7cm]{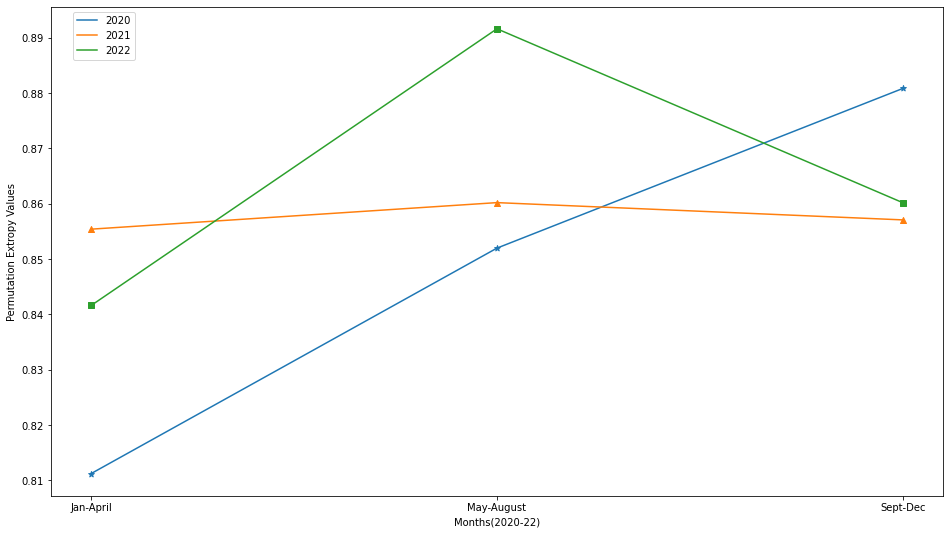} 
\caption{Comparison of permutation extropy values between different years.}
\label{fig:subim36}

\end{figure}
For analyzing the situation of covid during a year, we divide one year into three phases, the first is January to April, the second is May to August and the third is September to December. We plot the graph by taking different months along $X$ axis and permutation extropy values along $Y$ axis. For the year 2020, the covid was at its peak. Therefore the uncertainty in the number of death cases is more at that time. In Figure 17(b), the permutation extropy values during January-April are less because at that time the covid just started. After that during May-August, the uncertainty increases, hence the permutation extropy value increases. During September-December the situation was very worse, the fluctuation in the number of death cases was more. Therefore, the permutation extropy value during September-December is more in comparison to other phases. Similarly, we plot the graph for the year 2021, which can be shown in Figure 17(c). During January - April, there was a lockdown and there was not much fluctuation in the number of death cases that can be easily observed by taking the data of the death cases and comparing the data between any day and the previous day. The number of death cases almost oscillates around the same value. As the lockdown was open after this phase, which results in a large fluctuation in the number of death cases and hence more uncertainty. The permutation extropy value is high in comparison to other phases. After that, the uncertainty decreases slowly and hence the permutation extropy value is less in the duration September-December. After that, we plot the graph for the year 2022 similar to 2021. During January-April, the permutation extropy value is less. But during May-August, the uncertainty increases and hence the permutation extropy value is high in comparison to other phases. But after that during September-December, the uncertainty decreases and so the permutation extropy value decreases.

\section{Conclusions}
In these recent years, constant efforts are being made for quantifying the behavior of complex or chaotic systems more accurately by proposing different complex measures. In this paper, we propose a time series complexity measure, called as permutation extropy. We define the complexity measure by combining the concepts of extropy and permutation entropy which are two different complexity measures. We validate our complexity measure with the help of different chaotic maps like the logistic map, Henon map and Burger map. With the help of a bifurcation diagram, we compare permutation extropy and permutation entropy measures which accurately implies the correct behavior of the logistic map. We observe that for a particular value of $r$, the logistic map doesn't exhibit chaotic behavior which means the complexity measure should be less, this behavior is correctly captured by the measure permutation extropy but not in the case of permutation entropy. Similarly, with the bifurcation diagram of Henon diagram by varying parameters $a$ and $b$, we analyze the behavior of both permutation extropy and permutation entropy. But in the case of the Henon map, both the complexity measures behave similarly. While analyzing the behavior of both complexity measures that is permutation entropy and permutation extropy with the help of the bifurcation diagram of Burger map by the varying parameter $a$ and parameter $b$, we observe that both the measures behave similarly. Also by taking the historical data of NIFTY50, we analyze the behavior of the Indian stock market with the help of permutation extropy measure. It is very difficult to analyze the behavior of the market by plotting only time series plots of large volumes of data, hence by using our proposed complexity measure we can easily analyze all kinds of complex behaviors of the stock market. We also try to analyze the situation of covid in India by taking three-year data from WHO with the help of permutation extropy. We conclude that the proposed measure correctly describes and implies the situation during the covid period.

Although the permutation entropy is simple and can be first calculated for measuring the complexity of time series, our proposed measure is able to briefly describe the behavior of a complex system in some cases that we describe in this work while permutation entropy fails to do so. But further work can be done on developing different measures based on our proposed measure like weighted permutation extropy, multi-scale permutation extropy and multi-scale weighted permutation extropy and analyzing the behavior of these measures.

\end{document}